\documentstyle[amssymb,thmsa,sw20lart]{article}
\input{tcilatex}
\begin{document}

\title{Insufficient reason and entropy in quantum theory}
\author{Ariel Caticha \\
%EndAName
{\small Department of Physics, University at Albany-SUNY, Albany, NY 12222 }}
\date{}
\maketitle

\begin{abstract}
The objective of the consistent-amplitude approach to quantum theory has
been to justify the mathematical formalism on the basis of three main
assumptions: the first defines the subject matter, the second introduces
amplitudes as the tools for quantitative reasoning, and the third is an
interpretative rule that provides the link to the prediction of experimental
outcomes. In this work we introduce a natural and compelling fourth
assumption: if there is no reason to prefer one region of the configuration
space over another then they should be `weighted' equally. This is the last
ingredient necessary to introduce a unique inner product in the linear space
of wave functions. Thus, a form of the principle of insufficient reason is
implicit in the Hilbert inner product. Armed with the inner product we
obtain two results. First, we elaborate on an earlier proof of the Born
probability rule. The implicit appeal to insufficient reason shows that
quantum probabilities are not more objective than classical probabilities.
Previously we had argued that the consistent manipulation of amplitudes
leads to a linear time evolution; our second result is that time evolution
must also be unitary. The argument is straightforward and hinges on the
conservation of entropy. The only subtlety consists of defining the correct
entropy; it is the {\em array} entropy, not von Neumann's. After unitary
evolution has been established we proceed to introduce the useful notion of
observables and we explore how von Neumann's entropy can be linked to
Shannon's information theory. Finally, we discuss how various connections
among the postulates of quantum theory are made explicit within this
approach.
\end{abstract}

\section{Introduction}

Quantum theory is a set of rules for reasoning in situations where even
under optimal conditions the information available to predict the outcome of
an experiment may still turn out to be insufficient. This explains why the
notion of probability plays such a central role and immediately raises a
number of interesting questions.

One such question is whether these quantum probabilities differ in any
essential way from ordinary classical probabilities. It is sometimes argued
that there is an element of subjectivity in the nature of classical
probabilities that is not shared by quantum probabilities, that the latter
are totally objective because they are given by the Born rule, that is, by
the modulus squared of the wave function. One of the purposes of this paper
is to support the opposite point of view: we will argue that the
probabilities assigned using the Born rule are neither more nor less
subjective than say, the probability $1/6$ assigned to each face of a die
when there is no reason to favor one face over another. We will show that
there is a form of the principle of insufficient reason implicitly encoded
into the usual postulates of quantum theory.

A second question is the following. One would expect that if predictions are
to be made on the basis of insufficient information then quantities that
measure the amount of information, entropies, should play a central role 
\cite{Shannon48}\cite{Jaynes83}\cite{Kullback59}. Remarkably, one finds that
while the notion of entropy is indeed extremely useful, its use in
foundational issues has been very limited \cite{von Neumann55}\cite
{Deutsch83}\cite{Blankenbecler85}. Entropy is not mentioned in the
postulates; it is introduced later either to analyze quantum measurements or
in statistical mechanics where problems are sufficiently complicated that
clean deductive methods fail and one is forced to use dirtier inference
methods. A second purpose of this paper is to show that entropic arguments
are, in fact, implicit in the usual quantum postulates.

This paper is a continuation of previous work \cite{Caticha98a}\cite
{Caticha98b} in which quantum theory is formulated as the only consistent
way to manipulate amplitudes. In this consistent-amplitude quantum theory
(CAQT) amplitudes have a clear interpretation: they are tools for reasoning
that encode information about how complicated experimental setups are
related to those more elementary setups from which they were built. The
result of this approach is the standard quantum theory \cite{Dirac58}\cite
{von Neumann55}, in a form that is very close to Feynman's \cite{Feynman48}.

The objective of CAQT has been to justify the mathematical formalism on the
basis of rather general assumptions in the hope that this would not only
clarify the formal connections among the various postulates of quantum
theory but also illuminate the issue of how the formalism should be
interpreted. In this respect the traditional approach has been to set up the
formalism first and then try to find out what it all means. This problem of
attributing physical meaning to mathematical constructs is a notoriously
difficult one. So, instead of taking the standard quantum theory as
axiomatized by, say, von Neumann, and then, appending an interpretation to
it, the approach we take is to build the formalism and its interpretation
simultaneously.

In the brief summary of the CAQT given in section 2 three of the main
assumptions are explicitly stated. The first concerns the subject matter:
quantum theory is concerned with predicting the outcomes of experiments
performed with certain setups. The second introduces amplitudes as the tools
for quantitative reasoning, and the third is an interpretative rule that
provides the link between the mathematical formalism and the actual
prediction of experimental outcomes.

It is quite remarkable that although the interpretative rule does not in
itself involve probabilities it can be used to prove the Born statistical
`postulate' (or, better, the Born `rule') \cite{Finkelstein63} provided one
extra ingredient is added. The need for this fourth assumption arises
because the application of the interpretative rule requires a criterion to
quantify the change in amplitudes when setups are modified. In \cite
{Caticha98b} the criterion adopted was to use the Hilbert norm as the means
to measure the distance between wave functions. Such a technical assumption
without any obvious physical basis clearly detracts from the beauty and
cogency of the argument. In section 3 this blemish is corrected; we do not
remove the assumption, we just rewrite it in a form that is physically more
appealing and suggestive \cite{Caticha98c}. The main point is that the
components out of which setups are built, the filters, already supply us
with a notion of orthogonality and this takes us a long way towards defining
an inner product. Thus, instead of a strong and unnatural assumption about
the Hilbert norm we make a much weaker and more natural assumption about the
inner product.

The fourth assumption takes the form of a symmetry argument: if there is no
reason to prefer one region of the configuration space over another then
they should be weighted equally. The mere fact that some such assumption is
necessary already has interesting implications. The fact that a form of the
principle of insufficient reason is implicit in the Hilbert inner product
brings quantum probabilities closer to their classical counterparts. Quantum
probabilities are not more objective than classical probabilities. The
interpretation of quantum mechanics, just like that of other theories of
inference, is affected by the issue of what probabilities mean.

Once one finds that time evolution must be linear \cite{Caticha98a} the
obvious next question is whether it must also be unitary. These two issues
of linearity and unitarity are usually considered together (for a short
review see {\em e.g.} \cite{Jordan91}). A common explanation for the
unitarity of time evolution is that it guarantees the conservation of
probabilities. This is true but it is also irrelevant; that probabilities
should add up to one is true by definition \cite{Cox46}, and any
non-conservation of probabilities can always be trivially fixed by
reinterpreting $\left| \Psi \right| ^2$ as a relative probability rather
than the probability itself.

Another common explanation based on Wigner's theorem \cite{Wigner59} is also
found to be inadequate. The idea is to start with a quantum kinematics given
by a Hilbert space and deduce linear and unitary evolution from the
assumption that time evolution is a `symmetry' by which it is meant a
transformation that preserves orthogonality among states. The question, of
course, is why should time evolution be a `symmetry' in this technical
sense. In fact, when the assumption is relaxed one finds, as expected, that
the corresponding dynamics is non-unitary and irreversible \cite{Daniel82}.

This suggests yet another approach. It is a matter of definition that
entropy, as a measure of amount of information, is conserved whenever the
information available for the prediction of experimental outcomes is not
spoiled by the mere passage of time. The plan, then, is simple: impose
entropy conservation and from this deduce unitary time evolution. There is,
however, one remaining obstacle: one must identify the correct entropy. The
obvious candidate, von Neumann's entropy, fails. The problem is that the
interpretation, the very meaning of von Neumann's entropy, is derived in the
context of a linear quantum theory that is already assumed to be unitary 
\cite{von Neumann55}. Therefore, arguments based on von Neumann's entropy
are circular.

The argument we offer in section 4 is based on the idea of array entropy, a
concept that was briefly introduced by Jaynes \cite{Jaynes57} only to be
dismissed as an inadequate candidate for the entropy of a quantum system, a
quantity which he identified with von Neumann's entropy. From the point of
view of the CAQT, however, amplitudes and wave functions are assigned not
just to the system but to the whole experimental setup, and this turns the
array entropy into a legitimate entropy for our purpose. Its conservation
implies the conservation of the Hilbert norm and unitary evolution. As
claimed above, the notion of entropy plays an important role at the
foundations of quantum theory; it is implicit in the postulate that time
evolution is unitary.

Up to this point the discussion has been about experiments involving
idealized detectors localized at a given point in the configuration space.
In the traditional language the only observable measured is position. In
section 5 we address the issue of how observables other than position make
their appearance within the CAQT approach. We find that these observables
are useful concepts in that they facilitate the description of complex
experiments but, from our point of view, they are of only secondary
importance and play no role at the foundational level.

The prominence awarded within the CAQT to the concept of array entropy stems
partly from our choice of subject matter -- experimental setups rather than
quantum systems -- and partly from the fact that it is the array entropy
that provides the link between the Shannon information theory entropy and
von Neumann's entropy. The short discussion in section 6 shows two ways to
introduce von Neumann's entropy. This is an adaptation of the arguments of
Jaynes \cite{Jaynes57} and of Blankenbecler and Partovi \cite
{Blankenbecler85}.

We conclude in section 7 with a summary of our results and a discussion of
how various relations and connections among the postulates of quantum theory
are made explicit and clarified within the CAQT approach.

\section{The consistent-amplitude approach to quantum theory}

We proceed in several steps; effectively, each step consists of making an
assumption and then exploring its consequences. The first and most crucial
assumption is a decision about the subject matter. What problem is quantum
mechanics trying to solve? We choose a pragmatic, operational approach:
statements about a system are identified with those experimental setups
designed to test them \cite{Caticha98a}\cite{Caticha98b}. Our strategy is to
establish a network of relations among setups in the hope that information
about some setups might be helpful in making predictions about others. We
find that there are two basic kinds of relations among setups, which we call 
$and$ and $or$. These relations or operations represent our idealized
ability to build more complex setups out of simpler ones, either by placing
them in ``series'' or in ``parallel''.

Let us be more specific. To avoid irrelevant technical distractions we
consider a very simple system, a particle that lives on a discrete lattice
and has no spin or other internal structure. The generalization to more
complex configuration spaces should be straightforward. The simplest
experimental setup, denoted by $[x_f,x_i]$, consists of placing a source
that prepares the particle at a space-time point $x_i=(\vec{x}_i,t_i)$ and
placing a detector at $x_f=(\vec{x}_f,t_f)$. To test a more complex
statement such as ``the particle goes from $x_i$ to $x_1$ and from there to $%
x_f$,'' denoted by $[x_f,x_1,x_i]$, requires a more complex setup involving
an idealized device, a ``filter'' which prevents any motion from $x_i$ to $%
x_f$ except via the intermediate point $x_1$. This filter is some sort of
obstacle or screen that exists only at time $t_1$, blocking the particle
everywhere in space except for a small ``hole'' around $\vec{x}_1$. The
possibility of introducing many filters each with many holes leads to
allowed setups of the general form 
\begin{equation}
a=[x_f,s_N,s_{N-1},\ldots ,s_2,s_1,x_i]\text{ ,}
\end{equation}
where $s_n=(x_n,x_n^{\prime },x_n^{\prime \prime },\ldots )$ is a filter at
time $t_n$, intermediate between $t_i$ and $t_f$, with holes at $\vec{x}_n,%
\vec{x}_n^{\prime },\vec{x}_n^{\prime \prime },\ldots $

The first basic relation among setups, which we call $and$, arises when two
setups $a$ and $b$ are placed in immediate succession resulting in a third
setup denoted by $ab$. It is necessary that the destination point of the
earlier setup coincide with the source point of the later one, otherwise the
combined $ab$ is not allowed. The second relation, called $or$, arises from
the possibility of opening additional holes in any given filter.
Specifically, when (and {\em only} when) two setups $a^{\prime }$ and $%
a^{\prime \prime }$ are identical except on one single filter where none of
the holes of $a^{\prime }$ overlap any of the holes of $a^{\prime \prime }$,
then we may form a third setup $a$, denoted by $a^{\prime }\vee a^{\prime
\prime }$, which includes the holes of both $a^{\prime }$ and $a^{\prime
\prime }$. Provided the relevant setups are allowed the basic properties of $%
and$ and $or$ are quite obvious: $or$ is commutative, but $and$ is not; both 
$and$ and $or$ are associative, and finally, $and$ distributes over $or$. We
emphasize that these are physical rather than logical connectives. They
represent our idealized ability to construct more complex setups out of
simpler ones and they differ substantially from their Boolean and quantum
logic counterparts. In Boolean logic not only $and$ distributes over $or$
but $or$ also distributes over $and$ while in quantum logic propositions
refer to quantum properties at one time rather than to processes in time.

The identification of the $and/or$ relations, as well as their properties
(associativity, distributivity, etc.) is crucial to defining what kind of
setups we are talking about and therefore crucial to establishing the
subject of quantum mechanics. Thus, our first assumption is

\begin{itemize}
\item[{\bf A1.}]  The goal of quantum theory is to predict the outcomes of
experiments involving setups built from components connected through $and$
and $or$.
\end{itemize}

\noindent It is important to emphasize that this{\em \ }quantum theory
coincides with the standard Copenhagen quantum theory (see \cite{Stapp72}).
The contribution, at this point, has been to make explicit the relations $%
and/or$ which are normally implicit in the Feynman approach \cite{Feynman48}.

The next step involves an assumption about the existence of a quantitative
tool to handle these relations $and/or$:

\begin{itemize}
\item[{\bf A2.}]  A mathematical representation of $and/or$ is established
by the consistent assignment of a single complex number $\phi (a)$ to each
setup $a$ in such a way that relations among setups translate into relations
among the corresponding complex numbers.
\end{itemize}

\noindent What gives the theory its robustness, its uniqueness, is the
requirement that the assignment be consistent. If there are two different
ways to compute the single number $\phi (a)$ is assigned to setup $a$ the
two answers must agree. The remarkable consequence of this consistency
constraint is the `regraduation' theorem \cite{Caticha98b} that all such
representations are equivalent: changing representations involves mere
changes of variables. Thus, one can always `regraduate' $\phi (a)$ with a
function $\psi $ to switch to an equivalent and more convenient
representation, $\psi (a)\equiv \psi (\phi (a))$, in which $and$ and $or$
are respectively represented by multiplication and addition. Explicitly, $%
\psi \left( ab\right) =\psi \left( a\right) \psi \left( b\right) $ and $\psi
\left( a\vee a^{\prime }\right) =\psi \left( a\right) +\psi \left( a^{\prime
}\right) $. Anticipating the important role played by these conveniently
assigned complex numbers we call them by the suggestive name of
`amplitudes'. These amplitudes have a clear meaning, they are tools for
reasoning quantitatively and consistently about the relations $and/or$. For
an earlier derivation of the quantum sum and product rules see ref. \cite
{Tikochinsky88}. For comments on the possibility that such a representation
of $and/or$ might not exist see ref. \cite{Caticha98d}.

The observation that a single filter that is totally covered with holes is
equivalent to having no filter at all leads to the fundamental equation of
evolution. The idea is expressed by writing the relation among setups 
\begin{equation}
\lbrack x_f,x_i]=\bigvee_{\text{all}\,\,\vec{x}\,\text{at}%
\,t}([x_f,x_t][x_t,x_i])
\end{equation}
in terms of the corresponding amplitudes \cite{Feynman48}. Using the sum and
product rules, we get 
\begin{equation}
\psi (x_f,x_i)=\sum_{\text{all}\,\vec{x}\,\text{at}\,t}\psi (x_f,x_t)\,\psi
(x_t,x_i)\text{ .}  \label{motion1}
\end{equation}

Following Feynman \cite{Feynman48}, we introduce the wave function%
\index{wave function} $\Psi (%
\vec{x},t)$ as the means to represent those features of the setup prior to $%
t $ that are relevant to time evolution after $t$. Notice that there are
many possible combinations of starting points $x_i$ and of interactions
prior to the time $t$ that will result in identical evolution after time $t$%
. What these different possibilities have in common is that they all lead to 
\hspace{0pt}the same numerical value for the amplitude $\psi (x_t,x_i)$.
Therefore we set $\Psi (\vec{x},t)=\psi (x_t,x_i)$ and all reference to the
irrelevant starting point $x_i$ is omitted. The traditional language is that 
$\Psi $ describes the state of the particle at time $t$, that the effect of
the interactions was to prepare the particle in state $\Psi $. Now we see
that the word ``state'' just refers to a concise means of encoding
information about the preparation procedure prior to time $t$ that is
relevant for the evolution after time $t$.

The equation of evolution (\ref{motion1}) can then be written as

\begin{equation}
\Psi (\vec{x}_f,t_f)=\sum_{\text{all }\,\vec{x}\,\text{at}\,t}\psi (\vec{x}%
_f,t_f;\vec{x},t)\,\Psi (\vec{x},t)\text{,}  \label{motion}
\end{equation}
\newline
which is equivalent to a linear Schr\"{o}dinger equation as can easily be
seen \cite{Caticha98a}\cite{Caticha98b} by differentiating with respect to $%
t_f$ and evaluating at $t_f=t$. Thus, a quantum theory formulated in terms
of consistently assigned amplitudes must be linear; nonlinear modifications
of quantum mechanics must violate assumptions {\bf A1} or {\bf A2} else be
internally inconsistent \cite{Caticha98d}.

The question of how amplitudes or wave functions are used to predict the
outcomes of experiments is addressed through the time evolution equation (%
\ref{motion}). For example, suppose the preparation procedure is such that $%
\Psi (\vec{x},t)$ vanishes at a certain point $\vec{x}_0$. Then, according
to eq. (\ref{motion}), placing an obstacle at the single point $(\vec{x}%
_0,t) $ ({\it i.e.}, placing a filter at $t$ with holes everywhere except at 
$\vec{x}_0$) should have no effect on the subsequent evolution of $\Psi $.
Since relations among amplitudes are meant to reflect corresponding
relations among setups, it is natural to assume that the presence or absence
of the obstacle will have no effect on whether detection at $x_f$ occurs or
not. Therefore when $\Psi (\vec{x}_0,t)=0$ we predict that the particle will
not be detected at $(\vec{x}_0,t)$. This assumption can be generalized to
the following general interpretative rule:

\begin{itemize}
\item[{\bf A3.}]  If the introduction at time $t$ of a filter blocking those
components of the wave function characterized by a certain property ${\cal P}
$ has no effect on the future evolution of a particular wave function $\Psi
(t)$ then when the wave function happens to be $\Psi \left( t\right) $ the
property ${\cal P}$ will not be detected.
\end{itemize}

\noindent The application of this rule requires a means to quantify the
difference between wave functions before and after a filter. In ref. \cite
{Caticha98b} we showed how the interpretative rule {\bf A3} implies the Born
postulate provided this difference is measured by a Hilbert norm. In the
next section we justify this choice as being the uniquely natural one.

\section{The Hilbert inner product}

In order to justify the use of the Hilbert norm we show how the concepts of
distance and angle among states, that is an inner product, can be physically
motivated. The argument has three parts.

First, we note that wave functions form a linear space. To illustrate this
point suppose that $\Psi _1(\vec{x},t)=\psi (\vec{x},t;\vec{x}_1,t_0)$ is
the wave function at time $t$ of a particle that at time $t_0$ was prepared
at the point $\vec{x}_1$, and $\Psi _2(\vec{x},t)=\psi (\vec{x},t;\vec{x}%
_2,t_0)$ is the wave function at time $t$ of a particle that at time $t_0$
was prepared at the point $\vec{x}_2$. It is easy to prepare linear
superpositions of $\Psi _1(\vec{x},t)$ and $\Psi _2(\vec{x},t)$ by placing
the original source of the particle at an initial point $(\vec{x}_i,t_i)$
with $t_i$ earlier than $t_0$ and letting the particle evolve through a
filter at $t_0$ with two holes, one at $\vec{x}_1$ and the other at $\vec{x}%
_2$. Then the amplitude $\psi (\vec{x},t;\vec{x}_i,t_i)$ is 
\begin{equation}
\psi (\vec{x},t;\vec{x}_i,t_i)=\psi (\vec{x},t;\vec{x}_1,t_0)\psi (\vec{x}%
_1,t_0;\vec{x}_i,t_i)+\psi (\vec{x},t;\vec{x}_1,t_0)\psi (\vec{x}_1,t_0;\vec{%
x}_i,t_i)\text{ ,}
\end{equation}
and, in an obvious notation, the wave function at time $t$ is given by the
superposition 
\begin{equation}
\Psi (\vec{x},t)=\alpha \Psi _1(\vec{x},t)+\beta \Psi _2(\vec{x},t)\text{ .}
\end{equation}
Notice that the complex numbers $\alpha $ and $\beta $ can be changed at
will either by changing the starting point $(\vec{x}_i,t_i)$ or by modifying
the setup between $t_i$ and $t_0$ in any arbitrary way.

It is interesting that within the CAQT approach there is a deep connection
between the linearity of the space of wave functions and the linearity of
time evolution: they both follow from the same sum and product rules, and
ultimately, from consistency. In contrast, within the traditional approach 
\cite{Dirac58}\cite{von Neumann55} the two forms of linearity are seemingly
unrelated; the first is a kinematical feature while the second is dynamical.
In fact, attempts to formulate non linear variants of quantum theory give up
the second linearity, that of time evolution, while invariably preserving
the first \cite{deBroglie50}.

The second part of the argument is to point out that the basic components of
setups, the filters, already supply us with a concept of orthogonality
without invoking any additional assumptions.

The action of a filter $P$ at time $t$ with holes at a set of points $\vec{x}%
_p$ is to turn the wave function $\Psi (\vec{x})$ into the wave function 
\begin{equation}
P\Psi (\vec{x})=\sum_p\delta _{\vec{x},\vec{x}_p}\Psi (\vec{x})\text{ .}
\end{equation}
Since filters $P$ act as projectors, $P^2=P$, any given filter defines two
special classes of wave functions. One is the subspace of those wave
functions such as $\Psi _P\equiv P\Psi $ that are unaffected by the filter, $%
P\Psi _P=\Psi _P$. The other is the subspace of those wave functions that
are totally blocked by the filter, such as $\Psi _{1-P}\equiv (1-P)\Psi $,
for which $P\Psi _{1-P}=0$. We will say that these two subspaces are
orthogonal to each other.

Any wave function can be decomposed into orthogonal components, 
\begin{equation}
\Psi =P\Psi +(1-P)\Psi =\Psi _P+\Psi _{1-P}\text{ .}
\end{equation}
A particularly convenient expansion in orthogonal components is that defined
by a complete set of elementary filters. The filter $P_i$ is elementary if
it has a single hole at $\vec{x}_i$, it acts by multiplying $\Psi (\vec{x})$
by $\delta _{\vec{x},\vec{x}_i}$ and the set is complete if 
\begin{equation}
\sum_iP_i=1\text{ .}  \label{completeness}
\end{equation}
Then 
\begin{equation}
\Psi (\vec{x})=\sum\limits_iP_i\Psi (\vec{x})=\sum_iA_i\,\delta _{\vec{x},%
\vec{x}_i}\text{ .}
\end{equation}
where $A_i=\Psi (\vec{x}_i)$ and for $i\neq j$ the basis wave functions $%
\delta _{\vec{x},\vec{x}_i}$ and $\delta _{\vec{x},\vec{x}_j}$ are
orthogonal.

In the third and last step of our argument, as a matter of convenience, we
switch to the familiar Dirac notation. Instead of writing $\Psi (\vec{x})$
and $\delta _{\vec{x},\vec{x}_i}$ we shall write $|\Psi \rangle $ and $%
|i\rangle $, so that 
\begin{equation}
|\Psi \rangle =\sum_iA_i|i\rangle \text{ .}  \label{Psi}
\end{equation}
The question is what else, in addition to the notion of orthogonality
described above, is needed to determine a unique inner product. Recall that
an inner product satisfies three conditions:

\begin{description}
\item[(a)]  $\langle \Psi |\Psi \rangle \geqslant 0$ with $\langle \Psi
|\Psi \rangle =0$ if and only if $|\Psi \rangle =0$,

\item[(b)]  linearity in the second factor $\langle \Phi |\alpha _1\Psi
_1+\alpha _2\Psi _2\rangle =\alpha _1\langle \Phi |\Psi _1\rangle +\alpha
_2\langle \Phi |\Psi _2\rangle $ ,

\item[(c)]  antilinearity in the first factor, $\langle \Phi |\Psi \rangle
=\langle \Psi |\Phi \rangle ^{*}$.
\end{description}

\noindent Conditions (b) and (c) determine the product of the state $|\Phi
\rangle =\sum_jB_j|j\rangle $ with $|\Psi \rangle =\sum_iA_i|i\rangle $ in
terms of the product of $|j\rangle $ with $|i\rangle $, 
\begin{equation}
\langle \Phi |\Psi \rangle =\sum_iB_j^{*}A_i\langle j|i\rangle \text{ .}
\end{equation}
The orthogonality of the basis functions $\delta _{\vec{x},\vec{x}_i}$ is
easily encoded into the inner product, just set $\langle j|i\rangle =0$ for $%
i\neq j$. But the case $i=j$ remains undetermined, constrained only by
condition (a) to be real and positive. Clearly, an additional ingredient is
needed and to find it we reason as follows.

Suppose, that some prediction is made concerning the detection of the
particle at $\vec{x}_i$ when the state is $|\Psi \rangle $ (eq. \ref{Psi}).
Consider now another state $|\Psi ^{\prime }\rangle $ $=\sum_iA_i|i+k\rangle 
$ obtained from $|\Psi \rangle $ by a mere translation. What prediction
should we make concerning detection at $\vec{x}_{i+k}$? Since relations
among amplitudes are meant to reflect corresponding relations among setups,
it seems natural to assume that the latter prediction should coincide with
the former. As we show below this is achieved if we set $\langle i|i\rangle
=\langle i+k|i+k\rangle $, that is, we choose $\langle i|i\rangle $ equal to
a constant which, without losing generality, we may set equal to one.
Therefore, 
\begin{equation}
\langle i|j\rangle =\delta _{ij}\text{ .}
\end{equation}
This fixes a unique inner product 
\begin{equation}
\langle \Phi |\Psi \rangle =\sum_iB_i^{*}A_i\text{ ,}
\end{equation}
and yields the Hilbert norm 
\begin{equation}
\left\| \Psi \right\| ^2\equiv \langle \Psi |\Psi \rangle =\sum_i|A_i|^2%
\text{ .}
\end{equation}

Thus, we have arrived at the first main result of this paper: {\em the
principle of insufficient reason enters quantum theory through the inner
product}. Our assumption can in general be stated as

\begin{itemize}
\item[{\bf A4.}]  If there is no reason to prefer one region of
configuration space over another they should be assigned equal a priori
weight.
\end{itemize}

\noindent One should point out that the symmetry argument invoked here and
the usual symmetry arguments leading to conservation laws through Noether's
theorem are of a very different nature. The latter depends strongly on the
particular form of the Hamiltonian, on the dynamics; the former is totally
independent of the Hamiltonian.

The deduction of the Born statistical rule now proceeds as in ref. \cite
{Caticha98b}. Briefly the idea is as follows. We want to predict the outcome
of an experiment in which a detector is placed at a certain $\vec{x}_k$ when
the system is in state (\ref{Psi}). In \cite{Caticha98b} we showed that the
state for an ensemble of $N$ identically prepared, independent replicas of
our particle is the product 
\begin{equation}
|\Psi _N\rangle =\prod_{\alpha =1}^N|\Psi _\alpha \rangle =\sum_{i_1\ldots
i_N}A_{i_1}\ldots A_{i_N}\,|i_N\rangle ...|i_1\rangle \text{ .}
\end{equation}
Suppose that in the $N$-particle configuration space we place a special
filter, denoted by $P_{f,\varepsilon }^k$, the action of which is to block
all components of $|\Psi _N\rangle $ except those for which the fraction $%
n/N $ of replicas at $\vec{x}_k$ lies in the range from $f-\varepsilon $ to $%
f+\varepsilon $. The action of such a filter is represented by the projector 
\begin{equation}
P_{f,\varepsilon }^k=\sum_{n=(f-\varepsilon )N}^{(f+\varepsilon )N}P_n^k%
\text{ ,}
\end{equation}
where the $P_n^k$ are themselves projectors that select those components of $%
|\Psi _N\rangle $ for which the number of replicas at $\vec{x}_k$ is exactly 
$n$, 
\begin{equation}
P_n^k=\sum_{i_1\ldots i_N}|i_N\rangle ...|i_1\rangle \delta _{n,n_k}\langle
i_1|...\langle i_N|\qquad \text{where}\qquad n_k=\sum_{\alpha =1}^N\delta
_{k,i_\alpha }\text{ .}
\end{equation}

Next we prepare to apply the interpretative rule: we want to know whether
for large $N$ the state $P_{f,\varepsilon }^k|\Psi _N\rangle $ after the
filter differs or not from the state $|\Psi _N\rangle $ before the filter.
The distance between $P_{f,\varepsilon }^k|\Psi _N\rangle $ and $|\Psi
_N\rangle $, measured by the norm, 
\begin{equation}
\left\| P_{f,\varepsilon }^k|\Psi _N\rangle -|\Psi _N\rangle \right\| ^2%
\text{ ,}  \label{distance1}
\end{equation}
need not converge as $N\rightarrow \infty $, but the relative distance 
\begin{equation}
\frac{\left\| P_{f,\varepsilon }^k|\Psi _N\rangle -|\Psi _N\rangle \right\|
^2}{\left\| |\Psi _N\rangle \right\| ^2}\text{ }  \label{distance2}
\end{equation}
does. The calculation is straightforward \cite{Caticha98b}. We first
normalize $|\Psi \rangle $, 
\begin{equation}
\langle \Psi |\Psi \rangle =\,\sum_i|A_i|^2=1\text{ ,}
\end{equation}
so that $\langle \Psi _N|\Psi _N\rangle =1$ and the relative distance (\ref
{distance2}) coincides with (\ref{distance1}). The result is \newline
\begin{equation}
\left\| P_{f,\varepsilon }^k|\Psi _N\rangle -|\Psi _N\rangle \right\|
^2=1-\sum_{n=(f-\varepsilon )N}^{(f+\varepsilon )N}\binom Nn\left(
|A_k|^2\right) ^n\left( 1-|A_k|^2\right) ^{N-n}\text{ .}
\end{equation}
\newline
For large $N$ this binomial sum tends to the integral of a Gaussian with
mean $\overline{f}=|A_k|^2$ and variance $\sigma _N^2=\overline{f}(1-%
\overline{f})/N$. In the limit $N\rightarrow \infty $ this is more concisely
written as a $\delta $ function. Therefore \ 
\begin{equation}
\lim_{N\rightarrow \infty }\,\left\| P_{f,\varepsilon }^k|\Psi _N\rangle
-|\Psi _N\rangle \right\| ^2=1-\int_{f-\varepsilon }^{f+\varepsilon }\delta
\left( f^{\prime }-|A_k|^2\right) df^{\prime }.
\end{equation}
This shows that for large $N$ the filter $P_{f,\varepsilon }^k$ has a
negligible effect on the state $|\Psi _N\rangle $ provided $f$ lies in a
range $2\varepsilon $ about $|A_k|^2$. Therefore, according to the
interpretative rule {\bf A3}, the state $|\Psi _N\rangle $ does not contain
any fractions outside this range. On choosing stricter filters with $%
\varepsilon \rightarrow 0$, we conclude that detection at $\vec{x}_k$ will
occur for a fraction $|A_k|^2$ of the replicas and that it will not occur
for a fraction $1-|A_k|^2$. For any one of the {\em identical} individual
replicas however, there is no such certainty; the best one can do is to say
that detection will occur with a certain probability $\Pr (k)$. In order to
be consistent with the law of large numbers the assigned value must be, 
\begin{equation}
\Pr (k)=|A_k|^2\,.
\end{equation}

Theoretical arguments always involve idealizations which if taken literally
are clearly unrealistic. Some, such as the limit $N\rightarrow \infty $, are
obviously unphysical and yet routinely recognized as useful. But in other
cases legitimate doubts may arise. One may, for example, question whether in
invoking filters such as $P_{f,\varepsilon }^k$ acting on the $N$-particle
configuration space and selecting wave function components with a very
peculiar property ${\cal P}$ the idealizations are being pushed too far.
While recognizing that attempts to persuade all skeptics are doomed to fail,
perhaps the following two-dimensional example, borrowed from \cite{Farhi89},
may be of some help.

Consider the special case where the single particle state (\ref{Psi})
contains just two terms, say $|\Psi \rangle =A_1|1\rangle +A_2|2\rangle $.
This is analogous to a spin 1/2 particle. To pursue this analogy imagine the
individual spins of the $N$-particle ensemble are conveniently arranged in a
little crystal. This ensemble would have definite fractions of particles
with spin up (say, $|1\rangle $) without any individual spin being itself
definitely up or down. One way to determine this fraction consists of
sending the little crystal through a suitable Stern-Gerlach device to
measure its total spin by observing how it is deflected from the original
trajectory. The filter $P_{f,\varepsilon }^k$ is easy to visualize: it is
just a slit that allows passage when the deflection has the appropriate
value. According to the interpretative rule {\bf A3} the property ${\cal P}$
detected is that the fraction of particles with spin up is $|A_1|^2$. The
state of the ensemble $|\Psi _N\rangle $ as well as the state of each
individual spin remains unchanged in such a measurement.

But there is a second way to determine the definite value of the fraction of
spins up; it consists of detecting each individual spin and counting the
number with spin up. This second method will affect the state of the
ensemble as well as the state of each individual spin but it is a legitimate
way of detecting the same property ${\cal P}$ and whatever the result it
should agree with the previous method. Since there is a definite fraction of
spin up results and a definite fraction of spin down results for an
individual particle one can make no definite prediction. The best we can do
is assign a probability $|A_1|^2$ that the outcome will be spin up.

In our case we are concerned with position rather than spin but suitable
modifications are conceivable; perhaps the required `Stern-Gerlach' device
could directly measure the center of mass of the ensemble rather than its
total magnetization.

After this digression, let us return to assumption {\bf A4} and further
explore its implications. Suppose, for example, that the sites of the
discrete lattice on which the particle `moves' are unevenly spaced. Then
there is no reason to give equal weights to different $|i\rangle $'s. The
consequences of choosing a different normalization $\langle i|j\rangle
=w_i\delta _{ij}$ are easy to track down: the weighted inner product of $%
|\Phi \rangle =\sum_jB_j|j\rangle $ with $|\Psi \rangle =\sum_iA_i|i\rangle $
becomes 
\begin{equation}
\langle \Phi |\Psi \rangle =\sum_iw_iB_i^{*}A_i\text{ ,}
\end{equation}
the completeness relation (\ref{completeness}) becomes 
\begin{equation}
1=\sum\limits_iP_i=\sum\limits_iw_i^{-1}|i\rangle \,\langle i|\text{ ,}
\end{equation}
and the probability of detection at $\vec{x}_k$ would not be given by the
Born rule but rather by $\Pr (k)$ $=w_k|A_k|^2$.

An appealing but still arbitrary choice is to weight each cell of the
lattice by its own volume which we write as $w_i=g_i^{1/2}\Delta x$. This is
particularly interesting in the continuum limit $\Delta x\rightarrow 0$.
First, write the completeness relation (\ref{completeness}) as 
\begin{equation}
1=\sum\limits_ig_i^{1/2}\Delta x\frac{|i\rangle }{g_i^{1/2}\Delta x}\frac{%
\langle i|}{g_i^{1/2}\Delta x}\text{ .}
\end{equation}
On replacing $g_i^{1/2}\Delta x$ by $g^{1/2}dx$ and $(g_i^{1/2}\Delta
x)^{-1}|i\rangle $ by$\,|\vec{x}\rangle $ the new completeness condition
becomes 
\begin{equation}
\int g^{1/2}dx\,|\vec{x}\rangle \,\langle \vec{x}|=1\text{ .}
\end{equation}
Next, replace $\delta _{ij}/\Delta x$ by $\delta (\vec{x}-\vec{x}^{\prime })$
and the inner product $\langle i|j\rangle =g_i^{1/2}\Delta x\,\delta _{ij}$
becomes 
\begin{equation}
\langle \vec{x}|\vec{x}^{\prime }\rangle =g^{-1/2}\delta (\vec{x}-\vec{x}%
^{\prime })\text{ .}
\end{equation}
Furthermore, on replacing $A_i$ by $A(\vec{x})$, the state $|\Psi \rangle
=\sum\limits_iA_i\,|i\rangle $ becomes 
\begin{equation}
|\Psi \rangle =\int g^{1/2}dx\,\,A(\vec{x})\,|\vec{x}\rangle \text{ ,}
\end{equation}
and the Born rule $\Pr (i)=w_i|A_i|^2$ becomes 
\begin{equation}
\Pr (dx)=g^{1/2}dx\,|A(\vec{x})|^2\text{ .}
\end{equation}
As expected, $|A(\vec{x})|^2$ is the probability density. These results
apply to situations where the homogeneity of space is hidden by an
inconvenient choice of coordinates, and also to intrinsically inhomogeneous,
curved spaces.

We see that the Born rule follows, even in curved spaces, from giving the
same a priori weight, the same preference, to spatial volume elements that
are equal. This is a perhaps unexpected connection between quantum theory
and the geometry of space and one suspects that it is not accidental. It is
tempting to invert the logic and {\em assign} equal volumes to spatial
regions that are equally preferred. This would {\em explain} what a physical
volume is: just a measure of a priori preference. The full implications of
these remarks remain to be explored.

\section{Array entropy and unitary time evolution}

In a situation of optimal information everything that is relevant about the
experimental setup prior to time $t=0$ is known, then the wave function $%
\Psi (\vec{x},0)$ is known. But if less information is available perhaps the
best we can do is conclude that the actual preparation procedure was one
among several possibilities $\alpha =1,2,3,...$ each one with a certain
probability $p_{\alpha {}}$. For simplicity we initially assume these
possibilities form a discrete set. The usual linguistic trap is to say {\em %
the system} is in state $\Psi _{\alpha {}}(\vec{x},0)$ with probability $%
p_{\alpha {}}$, but it is better to say that {\em the preparation procedure}
is $\Psi _{\alpha {}}(\vec{x},0)$ with probability $p_{\alpha {}}$ \cite
{Peres84}. To this state of knowledge, which one may represent as a set of
weighted points in Hilbert space, and which Jaynes referred to as an array 
\cite{Jaynes57}\cite{footnote1}, one can associate an entropy, called the
array entropy 
\begin{equation}
S_A=-\sum_{\alpha {}}p_{\alpha {}}\log p_{\alpha {}}\text{ .}
\label{arrayentropy1}
\end{equation}
A valid objection to using this quantity as the entropy of the quantum
system is that if the $\Psi _{\alpha {}}(\vec{x},0)$ are not orthogonal then
the $p_{\alpha {}}$ are not the probabilities of mutually exclusive events.
When regarded as a property or an attribute of the quantum system the
various $\Psi _{\alpha {}}(\vec{x},0)$ need not, in fact, be mutually
exclusive; if $\langle \Psi _{{}\alpha }|\Psi _{\beta {}}\rangle \neq 0$,
knowing that the system is in $\Psi _{{}\alpha }(\vec{x},0)$ does not
exclude the possibility that it will be found in $\Psi _{{}\beta }(\vec{x}%
,0) $. However, if the $\Psi _{\alpha {}}(\vec{x},0)$ are attributes of the
preparation procedure then they are mutually exclusive because the
preparation devices are macroscopic! $S_A$ is a useful concept when
interpreted as the entropy of the whole setup and not as the entropy of the
quantum system.

The importance of this conceptual point cannot be overemphasized and a more
explicit illustration may clarify it further. Consider a spin $1/2$ particle
prepared either with spin along the $z$ direction or with spin along the $x$
direction. These states are not orthogonal and by `looking' at the particle
there is no sure way to tell which of the two alternatives holds, and yet
nothing prevents one from looking directly at the macroscopic Stern-Gerlach
magnets. This will reveal which of the two mutually exclusive orientations
was used without affecting the wave function. One can distinguish
non-orthogonal states by looking at the macroscopic devices that prepared
the system rather than by looking at the system itself.

Turning to the issue of time evolution, we consider situations where those
parts of the setup after time $0$ are known and no further uncertainty is
introduced. Under these conditions the points of the new array are shifted
from $\Psi _{\alpha {}}(\vec{x},0)$ to $\Psi _{\alpha {}}(\vec{x},t)$ but
their probabilities $p_{\alpha {}}$ and the corresponding array entropy $S_A$
remain unchanged.

The uncertainty discussed in the previous paragraphs led to a probability
distribution defined over a discrete array but, in general, we may have to
deal with a continuous array. This is of considerable significance for the
issue of time evolution.

The simplest continuous array is one dimensional, a weighted curve $C$ in
Hilbert space. We could consider higher dimensional arrays but this would
unnecessarily obscure the argument that follows. The
reparametrization-invariant entropy of this continuous array is \cite
{Jaynes63} 
\begin{equation}
S_A=-\int_Cd\alpha \,p(\alpha )\,\log \,\frac{p(\alpha )}{\ell (\alpha )}%
\text{ ,}  \label{arrayentropy2}
\end{equation}
where $p(\alpha )d\alpha $ is the probability that the preparation procedure
lies in the interval between $\alpha $ and $\alpha +d\alpha $ and $\ell
(\alpha )d\alpha $ is a measure of the distance in Hilbert space between $%
\Psi _{\alpha {}}(\vec{x},0)$ and $\Psi _{\alpha {}+d\alpha }(\vec{x},0)$.
As discussed in the last section the Hilbert norm is the uniquely natural
choice of distance, thus $\ell (\alpha )d\alpha =\left\| |\Psi _{\alpha
{}+d\alpha }\rangle -|\Psi _{\alpha {}{}}\rangle \right\| $.

Again we consider setups for which no further uncertainty is introduced
between times $0$ and $t$. We find that points $\Psi _{\alpha {}}(\vec{x},0)$
of the old line array at $t=0$ will move to points $\Psi _{\alpha {}}(\vec{x}%
,t)$ to form a new line array at time $t$. Since no information was lost
between times $0$ and $t$ we expect that, just as in the discrete case, the
probabilities $p(\alpha )d\alpha $ remain unchanged and the corresponding
array entropy $S_A$ is conserved. But entropy conservation, 
\begin{equation}
\frac{\partial S_A}{\partial t}=\int_Cd\alpha \,\,\frac{p(\alpha )}{\ell
(\alpha )}\,\frac{\partial \ell (\alpha )}{\partial t}=0\text{ ,}
\end{equation}
should hold for any curve $C$ and any function $p(\alpha )$, therefore 
\begin{equation}
\frac{\partial \ell (\alpha )}{\partial t}=0\text{ .}
\end{equation}
The conservation of the array entropy leads to the conservation of Hilbert
space distances. Since linear transformations that preserve the Hilbert norm
are called unitary we conclude that time evolution is given by a unitary
transformation. The Hamiltonian must be Hermitian; energy eigenvalues are
real.%
\index{unitary time evolution}

In the argument above it is implicit that

\begin{itemize}
\item[{\bf A}$^{\prime }${\bf 5.}]  The experimental setups about which we
wish to make predictions involve no loss of information.
\end{itemize}

\noindent This assumption is of a somewhat different nature than the
previous ones -- thus the prime. Since the objective of {\bf A}$^{\prime }$%
{\bf 5} is to specify more precisely what are the experimental setups we are
dealing with, {\bf A}$^{\prime }${\bf 5} is in effect contributing to define
the subject of quantum theory. It may, therefore, be more appropriate to
include {\bf A}$^{\prime }${\bf 5} as part of {\bf A1}. On the other hand,
one can also make the case that {\bf A}$^{\prime }${\bf 5} is already
implicit in {\bf A2}: it is only to those setups that have been optimally
specified that one can assign a single complex number. In any case, the
purpose of {\bf A}$^{\prime }${\bf 5} is to make explicit that in these
setups entropy must be conserved.

\section{Observables other than position}

The experiments we have discussed involve position detectors. One could say
we have only considered position `measurements', but this usage of the word
`measurement' requires some caution. The problem is that it suggests that
before the `measurement' the particle had a position, the value of which,
albeit unknown, was very definite. This is an assumption that need not and
should not be made; statements about whether the particle has a position or
not should be avoided. These statements are not identifiable with
experimental setups, and according to our assumption {\bf A1}, they are
foreign to the subject matter of CAQT; they are not even wrong, they are
meaningless. What has a definite position is the detector, not the particle 
\cite{footnote2}.

The issue we address next concerns other observables, how they are
`measured' and what role they play.

To build more complex detectors one can modify the setup just prior to the
final position detection at $%
\vec{x}_f$ by introducing, for example, additional magnetic fields or
diffraction gratings. Suppose that the setup prior to time $t$ prepares the
system in a certain state. After time $t$ the time evolution will in general
be given by eq. (\ref{motion}) but suppose that interactions between the
time $t$ and the time of detection $t_f$ are arranged in such a way that if
the wave function happened to be the function $\Phi _j(\vec{x},t)$ then at
the later time $t_f$ the new wave function $\phi _j(\vec{x},t_f)$ would
vanish everywhere except at $\vec{x}_j$,

\begin{equation}
\phi _j(\vec{x}_f,t_f)=\sum_{\text{all }\vec{x}}\psi (\vec{x}_f,t_f;\vec{x}%
,t)\,\Phi _j(\vec{x},t)=\delta _{\vec{x}_f,\vec{x}_j}.
\end{equation}
In this special case the particle would be detected at $\vec{x}_j$ with
certainty and we would say that ``at time $t_f$ the particle was found at $%
\vec{x}_j$''. Alternatively, we could describe this same result and convey
additional relevant information about the setup by saying that ``at time $t$
the particle was found in state $\Phi _j(\vec{x},t)$''. Thus, the latter
form of speech, although somewhat inappropriate, has the virtue of being
more informative.

The generalization is straightforward: arrange interactions so that each
state $\Phi _j(\vec{x},t)$ of a complete and orthogonal set is made to
evolve to a corresponding state $\phi _j(\vec{x},t_f)=\delta _{\vec{x},\vec{x%
}_j}$. The wave function at time $t$ can be expanded 
\begin{equation}
\Psi (\vec{x},t)=\sum_ja_j\,\Phi _j(\vec{x},t),  \label{expansion}
\end{equation}
and this evolves to

\begin{equation}
\Psi (\vec{x},t_f)=\sum_ja_j\delta _{\vec{x},\vec{x}_j}.
\end{equation}
Invoking the Born rule we interpret this as ``the probability that at time $%
t_f$ the particle is found at $\vec{x}_j$ is $\left| a_j\right| ^2$,'' or
alternatively, and somewhat inappropriately, that ``the probability that at
time $t$ the particle is found in state $\Phi _j(\vec{x},t)$ is $\left|
a_j\right| ^2$.''

What this particular complex detector `measures' is all observables of the
form $Q=\sum_nf_n|\Phi _n\rangle \langle \Phi _n|$ where the $f_n$ are
arbitrary scalars.

Notice that unitary evolution is a crucial requirement. In order for the
expansion (\ref{expansion}) to be unique the states $\Phi _n(\vec{x},t)$
must form a complete and orthogonal set which itself must evolve to the also
orthogonal set of $\phi _n(\vec{x},t_f)=\delta _{\vec{x},\vec{x}_n}$. The
orthogonality must be preserved. One cannot introduce this notion of
observables until after the issue of unitary time evolution has been settled.

Notice also that it is not necessary that the operator $Q$ have real
eigenvalues, but it is necessary that its eigenvectors $|\Phi _n\rangle $ be
orthogonal. This means that the Hermitian and anti-Hermitian parts of $Q$
must be simultaneously diagonalizable. Thus, while the observable $Q$ does
not have to be Hermitian ($Q=Q^{\dagger }$) it must certainly be {\em normal}%
, that is $QQ^{\dagger }=Q^{\dagger }Q$.

It is amusing to reflect that if the sentence ``at time $t$ a particle has
momentum $\vec{p}$'' is used only as a linguistic shortcut that conveys the
information that the wave function assigned to the setup prior to time $t$
was $exp(i\vec{p}\cdot \vec{x}/\hbar )$ then, strictly speaking, there is no
such thing as the momentum of the particle. The point is that wave functions
attach to setups and not to particles; whatever $\vec{p}$ is, it is not a
property of the particle by itself, but of the whole setup.

\section{von Neumann's entropy}

We saw that the array entropy (\ref{arrayentropy1}) is an acceptable measure
of uncertainty provided it is associated with the whole experimental setup
rather than the quantum system by itself. This interpretation hinged on the
fact that preparations are made using macroscopic devices with definite
attributes that are in principle distinguishable and mutually exclusive even
when the corresponding wave functions $\Psi _\alpha $ are not orthogonal.

But suppose that for some unspecified reason the part of the experimental
setup responsible for the preparation is not directly accessible to
observation and we can only look at the detectors themselves. This is what
happens when the actual purpose of the experiment is to obtain information
about the preparation procedure. Many, maybe most experiments are of this
kind. We can, for example, detect photons to obtain information about how
they were originally prepared in a distant star, and thereby we learn about
the star; or we can detect photons to find how they were prepared at the
other end of a communication channel, and thereby we receive a message. In
these cases a more useful, more relevant entropy might be one that measures
the uncertainty about how the detectors will respond.

Consider measuring an arbitrary observable $Q=\sum_nf_n|\Phi _n\rangle
\langle \Phi _n|$ in a situation where the preparation procedure is
uncertain. If the wave function is $\Psi _{\alpha {}}(\vec{x},0)$ with
probability $p_{\alpha {}}$ the probability that the system is detected in
state $|\Phi _n\rangle $ is 
\begin{equation}
p_n^Q=\sum_\alpha p_\alpha \left| \langle \Phi _n|\Psi _\alpha \rangle
\right| ^2=\langle \Phi _n|\rho |\Phi _n\rangle ,  \label{pnQ}
\end{equation}
where $\rho $ is the density operator 
\begin{equation}
\rho =\sum_\alpha p_\alpha |\Psi _\alpha \rangle \langle \Psi _\alpha |.
\end{equation}
Thus, knowledge of $\rho $ allows one to compute the probability of all
experimental outcomes. An important implication of this result is that if
all we can observe are the detectors then two different arrays with the same
density operator $\rho $ are indistinguishable; they yield experimental
outcomes that are statistically identical no matter what experiment is
performed. To distinguish among such arrays requires information which, in
practice, is just not available. A second important feature is that since $%
\rho $ is Hermitian it can be diagonalized, {\em i.e.}, 
\begin{equation}
\rho =\sum_\beta w_\beta |w_\beta \rangle \langle w_\beta |\text{ ,}
\end{equation}
where $\langle w_\beta |w_\gamma \rangle =\delta _{\beta \gamma }$.
Therefore, the set of all arrays with the same $\rho $ includes an array
that is orthogonal.

The von Neumann entropy can now be introduced in either of two ways. First,
we note that two different arrays with the same $\rho $ need not have the
same array entropy. What is remarkable is that even though for one array one
might have a higher uncertainty about the preparation procedure this will
not diminish our ability to predict the response of the detectors. As far as
the detectors are concerned the additional uncertainty was irrelevant. The 
{\em relevant} uncertainty of all these arrays with the same $\rho $ is the
minimum value that the array entropy can attain. It can be shown that the
minimizing array is the orthogonal one \cite{Jaynes57} (see also \cite
{Wootters93}). The corresponding entropy is von Neumann's 
\begin{equation}
S_{vN}(\rho )=\left. \min_AS_A\right| _\rho =-\sum_\beta w_\beta \log
w_\beta =-\func{Tr}\rho \log \rho \text{ ,}
\end{equation}

Notice that one cannot use the von Neumann entropy introduced in this first
way to argue that time evolution must be unitary. If no information is
dissipated one can reasonably expect that the array entropy of an array at
time $t_1$ should coincide with the array entropy at a later time $t_2$, but
there is no reason to expect that the {\em relevant} part of these
uncertainties should also coincide. In other words, a priori there is no
reason to assume that it is the orthogonal array at $t_1$ that evolves into
the orthogonal array at $t_2$.

A second way to introduce von Neumann's entropy is to focus attention
directly on the response of the detectors. The uncertainty about which
detector will fire when the observable $Q$ is being measured is given by the
so called {\em measurement} entropy 
\begin{equation}
S(\rho |Q)=-\sum_{n{}}p_n^Q\log p_n^Q\text{ .}
\end{equation}
with $p_n^Q$ given by (\ref{pnQ}). Notice that even if we have optimal
information about the preparation procedure, that is, even if $\rho $
represents a pure state, the measurement entropy need not vanish -- there
remains the uncertainty introduced by the measurement itself which is given
by the Born rule probabilities. This indicates that $S(\rho |Q)$ receives
contributions from both the uncertainty in the preparation procedure and
from the measurement itself. Naturally, the latter will depend on the choice
of the observable $Q$. If one seeks a measure of the uncertainty in the
preparation procedure one should choose that $Q$ which makes the least
contribution to $S(\rho |Q)$. The desired observable is $\rho $ itself \cite
{Blankenbecler85} and the corresponding entropy is von Neumann's, 
\begin{equation}
S_{vN}(\rho )={}\min_QS(\rho |Q)=S(\rho |\rho )=-\func{Tr}\rho \log \rho .
\end{equation}
Notice, again, that one cannot use the von Neumann entropy introduced in
this second way to argue that time evolution must be unitary. The problem is
that, as discussed in the previous section, the possibility of measuring
arbitrary observables $Q$ can only be established after the issue of unitary
time evolution has been settled.

To summarize, whichever way one chooses to introduce it, von Neumann's
entropy represents that component of the uncertainty in the preparation
procedures that is relevant to the response of the detectors.

\section{Final remarks}

The main goal of the CAQT is to justify the formalism of quantum theory on
the basis of rather general assumptions. An important by-product is that it
has revealed interesting connections among the various postulates of quantum
theory. To illustrate this point and, in this context, summarize our main
results, consider the following standard set of postulates:

\begin{description}
\item[{\bf P1}]  The states of a quantum system are represented by elements $%
|\psi \rangle $ in a linear space ({\bf P1a}) with an inner product ({\bf P2a%
}), {\em i.e.}, the $|\psi \rangle $ are vectors in a Hilbert space.

\item[{\bf P2}]  The time evolution $|\psi (t)\rangle =U(t)$ $|\psi
(0)\rangle $ is given by an operator $U(t)$ which is both linear ({\bf P2a})
and unitary ({\bf P2b}).

\item[{\bf P3}]  Every observable ${\cal A}$ is represented by a Hermitian
operator $A$. The outcome of a measurement of observable ${\cal A}$ is one
of the eigenvalues $a$ of the corresponding operator $A$, $A|a\rangle
=a|a\rangle $.

\item[{\bf P4}]  The Born postulate: the probability that the measurement of 
${\cal A}$ in a system in the normalized state $|\psi \rangle $ yields the
eigenvalue $a$ is given by $|\langle a|\psi \rangle |$ $^2$.

\item[{\bf P5}]  The projection postulate: after a measurement that yields
the eigenvalue $a$ the system is left in the eigenstate $|a\rangle $.
\end{description}

Consider first a possible connection between {\bf P1} and {\bf P2}. The idea
that the wave function is just a way to codify whatever information is
relevant for the purpose of making predictions about the future implies that
an adequate specification of the state will necessarily depend on the nature
of the laws ruling time evolution. Conversely, deciding on a law of time
evolution will depend on what it is that is evolving. But this connection
between {\bf P1} and {\bf P2} is not explicit in the usual approach. For
example, both postulates {\bf P1a} and {\bf P2a} refer to linearity, but
these seem to be unrelated, independent linearities. It appears possible to
give up the dynamical linearity in {\bf P2a} while preserving the
kinematical linearity in {\bf P1a}. In the traditional approaches to quantum
mechanics the kinematical aspects of the theory are kept isolated from the
dynamical ones. In contrast, within the CAQT approach kinematics and
dynamics are unified into a single structure and, in particular, there is
only one linearity, which follows from the consistency constraint in the
form of the sum and product rules. The resulting formalism is more rigid,
more robust; small modifications are not tolerated.

The remaining postulates {\bf P3}, {\bf P4} and {\bf P5} deal with
observations and measurements. Since these physical processes are themselves
ruled by {\bf P1} and {\bf P2}, it should be the case that the first two
postulates already have a lot to say about what is and is not observable and
what the allowed outcomes of measurements should be; parts of {\bf P3}, {\bf %
P4} and {\bf P5} are redundant. We find that {\bf P3} and {\bf P5} are
redundant except those aspects that refer to experiments involving position
detectors. Other observables are useful and convenient but not crucial. For
these observables {\bf P3} makes no contribution beyond what is already
contained in {\bf P1} and {\bf P2}.

We have also found that unitary time evolution {\bf P2b} and the Born
probability rule {\bf P4} are linked in yet another way, they both follow
from the Hilbert inner product {\bf P1b} which is itself a consequence of a
form the Principle of Insufficient Reason embodied in our assumption {\bf A4}%
.

The Born probability rule {\bf P4} is replaced by a milder and more
compelling assumption, the general interpretative rule {\bf A3} which does
not mention probabilities. From the point of view of the CAQT\ indeterminism
arises as a consequence of our assumption {\bf A2} that a {\em single}
complex number provides an optimal means of codifying information about a
setup and this information, while optimal, is definitely not sufficient. At
this point it is still an open question whether more information could be
codified into a single `larger' mathematical object (say, a matrix)
satisfying the associativity and distributivity requirements \cite
{Finkelstein62}\cite{Rodriguez98}. In any case the mystery remains: why
complex numbers? Our assumption {\bf A4} has made explicit what is an
intriguing and perhaps unexpected connection between the Hilbert inner
product and spatial measures of volume. Perhaps the reason for complex
numbers will be found in the geometry of space.

Lastly, we comment on the pragmatic interpretation implicit in the CAQT
approach. The interpretation and the formalism are irremediably entangled
form the very beginning. Already from assumption {\bf A1} the CAQT will make
no attempt to offer a model, a description of an underlying objective
reality (the existence of which is, however, never doubted). Depending on
one's prejudices this may either be totally unsatisfactory or it may be
highly desirable. It incorporates the deep insight of Bohr and Heisenberg
that the subject of physics is one step removed from reality. Physics does
not model reality itself, it models information about reality.

The interpretation of the wave function follows from assumption {\bf A2}:
amplitudes and wave functions are tools for reasoning. It is interesting
that since $\Psi (t)$ refers to a certain experiment and $\Psi (t+\delta t)$
to another, different, albeit closely related experiment, it is
inappropriate to say that $\Psi (t)$ has `moved' to $\Psi (t+\delta t)$.
Wave functions `evolve' in a sense that is somewhat analogous to the changes
between successive frames in a film; there may be an illusion but there is
no real motion. Thus the Schr\"{o}dinger equation is not an equation of
motion, it is an equation of evolution. Wave functions do not move and
consequently they do not collapse either.

While this pragmatic interpretation coincides in its most crucial aspects
with the Copenhagen interpretation \cite{Stapp72} there are some
differences. There is no need to invoke complementary features for the
description of the quantum system because the CAQT never attempts such a
description in the first place. The doctrine of complementarity is not
needed, and in fact it would represent a step in the wrong direction, a
half-hearted attempt to peek behind the curtain and at least partially
describe `what is really going on'. Similarly, since quantities are not
associated to the system by itself there is no need to assert that the
values of certain quantities associated to the system are created by acts of
measurement. In fact, there is no `measurement.'

{\bf Acknowledgments-} I am indebted to C. Rodr\'{\i }guez for many valuable
discussions. Correspondence with J. Hartle and L. Schulmann on the issue of
the Hilbert norm is also gratefully acknowledged.


\begin{thebibliography}{99}
\bibitem{footnote1}  If the states are normalized the points of the array
lie on the surface of a unit sphere, but normalization is not necessary for
our argument.
\end{thebibliography}
\end{document}